\documentclass[english,aps,manuscript]{revtex4}
\usepackage[T1]{fontenc}
\usepackage{hyperref}
\usepackage[latin9]{inputenc}
\usepackage[letterpaper]{geometry}
\usepackage{braket}
\usepackage{setspace}
\usepackage{amssymb,amsmath}
\makeatletter
\@ifundefined{textcolor}{}
{%
 \definecolor{BLACK}{gray}{0}
 \definecolor{WHITE}{gray}{1}
 \definecolor{RED}{rgb}{1,0,0}
 \definecolor{GREEN}{rgb}{0,1,0}
 \definecolor{BLUE}{rgb}{0,0,1}
 \definecolor{CYAN}{cmyk}{1,0,0,0}
 \definecolor{MAGENTA}{cmyk}{0,1,0,0}
 \definecolor{YELLOW}{cmyk}{0,0,1,0}
 }

\makeatother

\usepackage{babel}

\begin{document}

\title{SECURITY SIGNIFICANCE OF THE TRACE DISTANCE CRITERION IN QUANTUM KEY DISTRIBUTION}

\author{Horace P. Yuen }

\email{yuen@eecs.northwestern.edu }

\affiliation{Department of Electrical Engineering and Computer Science }

\affiliation{Department of Physics and Astronomy, Northwestern University, Evanston,
IL 60208}
\begin{abstract}
The security significance of  the trace distance security criterion $d$ is analyzed in terms of operational probabilities of an attacker's success in identifying different subsets of the generated key, both during the key generation process and when the key is used in one-time pad data encryption under known-plaintext attacks. The difference between Eve's sequence error rate and bit error rate is brought out. It is shown with counter-examples that the strong security claim maintained in the literature is incorrect. Other than the whole key error rates that can be quantified at the levels $d^{1/3}$ and $d^{1/4}$  which are much worse than $d$ itself, the attacker's success probabilities in estimating various subsets of the key and in known-plaintext attacks are yet to be quantified from $d$ if possible. It is demonstrated in realistic numerical examples of concrete protocols that drastic breach of security cannot yet be ruled out.
\end{abstract}
\maketitle

\section{INTRODUCTION}\label{sec:intro}
   In quantum key distribution (QKD) there have been many proofs offered on the ``unconditional security'' of various protocols of the BB84 variety. For a recent review see \cite{1}. Security is a serious matter which cannot be established experimentally in contrast to most problems in science and technology, if only because the possible different attacks are unlimited. A systematic examination of fundamental security analysis, as opposed to practical imperfections, will be offered of which this paper constitutes the first and is concerned with security criteria, their {\it empirical meanings} and the {\it quantitative levels} needed for meaningful security. Without them the term ``unconditional'' or whatever security would be empty and in fact misleading. It will be shown that, in contrast to the prevalent claims in the literature, QKD security has not yet been properly quantified.

Until 2004-2005 and often till recently, the security criterion adopted is the attacker Eve's quantum accessible information ($I_{ac}$) on the generated key $K$, which is the maximum mutual information Eve has on $K$ from a measurement on her probe she may set on the quantum signals transmitted during the key generation process.  Security of $K$ during key generation before it is actually used is called ``raw security''\cite{2}, to distinguish it from ``composition security'' when $K$ is actually used in an application for which Eve may possess additional information related to $K$. In particular, when $K$ is used for data encryption, part of $K$ may be known to Eve in a known-plaintext attack (KPA) to help her get at the rest of $K$ and thus the rest of the encrypted data. KPA can be simply represented when $K$ is used in the one-time pad format, in which some bits among the $n$-bit $K$ are known to Eve. KPA security is necessary for security because that is the major weakness of conventional symmetric-key ciphers QKD purports to overcome. Indeed, there is otherwise no need for QKD since its achievable raw security is worse in a precise sense \cite{2} than that of conventional ciphers in which the key is typically totally hidden by uniform random data. The security under discussion is information theoretic and symmetric key cipher, not RSA,  is the proper comparison with QKD since a shared secret key is needed for both in an information theoretic context.

It was claimed in \cite{3} that an exponentially small (in $n$) $I_{ac}$ would guarantee universal composition security and that such is the case obtained in prior security proofs, while suggesting also a trace distance criterion, often denoted by ``$d$'' which we adopt, is a better criterion to work with. In \cite{4} it was shown from quantum information locking that mere exponentially small $I_{ac}$ does not imply KPA security. The small leak in \cite{4} is enlarged to a ``spectacular failure'' of the $I_{ac}$ criterion in \cite{5}. At present, $d$ is the only basis of QKD unconditional security claim including the use of privacy amplification \cite{1,6,7}.

However, the precise security claim associated with $d$ has not been quantitatively spelled out while various paraphrases in words without mathematical expression are given for very strong security claims on behalf of $d$. In \cite{2} the errors of some such interpretations
are pointed out, but there are other possible interpretations that appear to support such claims which still persist. To settle this fundamental security issue, in the following we will examine all such interpretations one may infer from the various wordings in the literature and give them precise mathematical expressions. They are found to be either incorrect or too weak to support the claimed security. Generally, Eve's optimal sequence error rate on $K$ is only bounded by $d^{1/3}$ and her bit error rate is bounded from $\frac{1}{2}$ by $d^{1/4}$, while her error rates on subsets of $K$ are not quantified. In the process, we also identify the empirically meaningful criteria of security that guarantee fundamental security. It will be demonstrated that disastrous breach of security is not ruled out with practically achievable levels of $d$ from concrete protocols, even though that is very unlikely with probability bounded by $d$ under the prevalent incorrect interpretation.

\section{THE CRITERION $d$ AND ITS INTERPRETATIONS }

During key generation Eve sets her probe and the protocol goes forward.  After privacy amplification the final key $K$ is generated with corresponding ``prior probability" $p(k)$ and probe state $\rho^k_E$ on each $k$. Let

\begin{equation}
\rho = \sum_k p(k)\ket{k}\bra{k}
\end{equation}

\noindent for $N$ orthonormal $\ket{k}$'s in space $\mathcal H_K$, $N=2^n$. Let

\begin{equation}
\rho_E = \sum_k p(k) \rho_E^k
\end{equation}

\begin{equation}
\rho_{KE} = \sum_k p(k)\ket{k}\bra{k}\otimes \rho_E^k
\end{equation}

\noindent The criterion $d$ is defined to be

\begin{equation}
d \equiv \frac{1}{2}\parallel\rho_{KE}-\rho_U\otimes\rho_E\parallel_1
\end{equation}

\noindent where $\rho_U$ is given by (1) with $p(k)=U(k)$ for the uniform random variable $U$. It can be readily shown (similar to Lemma 2 in \cite{6}) that

 \begin{equation}
d = \frac{1}{2}\sum_k \parallel p(k)\rho_E^k-\frac{1}{N}\rho_E\parallel_1
\end{equation}
A key $K$ with $d\leq\epsilon$ is called ``$\epsilon$-secure'', as it has been forced by privacy amplification to be $\epsilon$-close to $U$. But what is the operational meaning of $d\leq \epsilon$?

The major interpretation that has been given to $d\leq\epsilon$ which has lent it a strong security interpretation has two slightly different versions, to be called (i) and (ii) in the following:

\begin{enumerate}
\item[(i)] ``The real and the ideal setting can be considered to be identical with probability at least 1-$\epsilon$." \cite{4,6}
\item[(ii)]  The parameter $\epsilon$ can be understood as the ``maximum failure probability'' \cite{1, 8,9} of the real protocol, ``where `failure' means that `something went wrong', e.g., that an adversary might have gained some information on $K$'' \cite{8}.
\end{enumerate}

In \cite{2} the interpretation (i) is shown to be incorrect in that the real and the ideal setting are actually different with probability $1$ from the given mathematical representation. We now give a precise mathematical representation of (ii), namely, Eve's real performance $\mathcal{P}_r$ (the bigger the better for her) in an attack of any kind can be better than that of $\mathcal{P}_i$, the performance of the ideal uniform $K$ case, by a probability no more than $\epsilon$, i.e.,
                                                                                                                                                                 \begin{equation}
\Pr [\mathcal{P}_r >\mathcal{P}_i]\leq \epsilon
\end{equation}

\noindent The probability in this case is obtained when Eve makes a measurement, as opposed to (i) which involves the setting before any measurement. This intended probability interpretation of $\epsilon$ is also utilized explicitly in \cite{9} to derive composition security under $d$.

The justification for (ii) is the same as for (i), namely the interpretation of Lemma 1 in \cite{6} that the variational distance $v(P,Q), 0\leq v \leq 1$, between two distributions $P$ and $Q$ on the same sample space ``can be interpreted as the probability that two random experiments described by $P$ and $Q$, respectively, are different''. This interpretation is derived from Lemma 1 that asserts the {\it existence} of a joint distribution which gives marginals $P$ and $Q$ and for which the results of the two random experiments differ with just probability $v(P,Q)$. It was pointed out \cite{2,10} that there is no reason to expect there is such a joint distribution in force. However,  the contrary belief is still widespread.

Thus, two more reasons why this would not work are now given, to be followed by specific counter-examples. First, even if such joint  distribution is in force it is {\it not} represented in the mathematical formulation, which just treats $P$ and $Q$ as independent. Furthermore, if it is introduced explicitly then whether the measurement result is close to that from $U$ is irrelevant since Eve can learn about $K$ through the joint distribution, i.e., the claimed conclusion still does not follow. In fact counter-example would result, classically or quantum mechanically, whenever knowing parts of $K$ reveals the rest for certainty such as in the case of information locking.

For an explicit simple example, consider a two-bit $K$ with $k = k_1 k_2$  for two bits  $k_1$ and $k_2$ with uniform a priori probability. Let $\ket{i}, i\in \overline {1-4}$ be the four BB84 states on a qubit with $\Braket{1|3}=\Braket{2|4}=0$. Let the states be
\begin{equation}
\begin{split}
\rho_E^{11}=\ket{1}\ket{1}, {~~~~~~~~~}\rho_E^{10}=\ket{1}\ket{3}\\
\rho_E^{01}=\ket{1}\ket{2}, {~~~~~~~~~}\rho_E^{00}=\ket{1}\ket{4}
\end{split}
\end{equation}

\noindent Thus, $k_2$ is locked into the second qubit through $k_1$ and is unlocked by measuring on the 1-3 or 2-4 basis given the knowledge of $k_1$. The protocol then fails for sure but $d$ is not given by 1, which never happens in any protocol since that requires $\rho_E^k$ and $\rho_E$ to be orthogonal.

To give a raw security example with large ``failure probability'',  consider the distribution upon a measurement result with $P_i=\frac{1+\epsilon}{N}$ for $i\in \overline{1-\frac{N}{2}}$ and $P_i=\frac{1-\epsilon}{N}$ for $i\in \overline{(\frac{N}{2}+1)-N}$ so that $v(P,U)=\epsilon$. Then Eve gains ``information'' compared to the ideal case with probability $\frac{1}{2}$, not $\epsilon$. What happens, of course, is that no ``joint distribution'' other than the product one is included in the mathematical formulation, and conceptually $\epsilon$ is {\it not} an event probability though it may be the difference of two event probabilities. Although the information gained in this counter-example is small, it clearly shows that variational distance is not the probability that information is leaked.

There is another interpretation:

\begin{enumerate}
\item[(iii)] ``Distinguishability advantage'' between the real and the ideal protocols is bounded by $\epsilon$ \cite{3}.
\end{enumerate}

\noindent The distinguishability advantage that is referred to here is the trace distance (4) between the ``real'' and the ``ideal'' situations. From binary quantum decision \cite{11}   the optimum probability of success is
\begin{equation}
P_c = \frac{1}{2} + ||p_o\rho_o - p_1\rho_1||_1
\end{equation}

\noindent for two states $\rho_0$ and $\rho_1$ with a priori probability $p_0, p_1=1-p_0$. Clearly, Eve does not make such binary decision if only because $\mathcal H_K$ is not accessible to her. Indeed if it is, she can just measure on it to identify $K$. The entanglement form (4) is misleading even in the classical limit, and the form (5) on Eve's probe alone is appropriate for interpretation.

Equ (5) shows that each unnormalized trace distance is bounded by $\epsilon$ and thus the trace distance between two different $\rho_E^k$ is bounded by $4\epsilon$ from the triangular inequality,
                                                                                                                                                               \begin{equation}
\parallel p(k)\rho_E^k-p(k^\prime)\rho_E^{k^\prime}\parallel_1 \leq 4\epsilon
\end{equation}

\noindent Since $p(k)$ is the distribution of $K$ after privacy amplification, it is generally nonuniform with correlation between bits. It is {\it not} possible to draw the conclusion from (8)-(9) that $k$ and $k^\prime$ cannot be distinguished with probability better than $P_c = \frac{1}{2} + 2\epsilon$ even after renormalization of $p(k)$ and $p(k^\prime)$. Also it is only possible to conclude from (5) that
\begin{equation}
\parallel p(k)\rho_E^k-\frac{1}{N}\rho_E \parallel_1 \leq 2\epsilon
\end{equation}

\noindent and not the ``distinguishability advantage'' interpretation $\parallel \rho_E^k-\rho_E \parallel_1 \leq 2\epsilon$. This is so even after the application of Markov Inequality to obtain an individual guarantee at a level $\epsilon^\prime > \epsilon$.

Interpretation (iii) is not relevant, certainly not sufficient, for general security which involves $M$-ary decisions by Eve for $M=2$ to $M=2^n$. Note that even classically, the $M$-ary performance could only improve for Eve with decreasing $M$, on the absolute level at least. A level $\epsilon$ may appear adequate for bit decision compared to $\frac{1}{2}$, but is in fact very inadequate for $M$-ary decision as compared to $\frac{1}{M}$. In the information security literature a ``semantic security" binary decision condition similar to (7) has been obtained for the ``bounded storage model" \cite{12}  for ``message'' security, with $K$ and any $\tilde K \subset K$ being possible messages for comparison with QKD. Such condition is not strong enough for just raw security as just indicated, apart from KPA security.   However, in \cite{12} the security guarantee is much stronger than (9) from $d$ because it is valid for every pair of $k$ and $k^\prime$ as well as any $\tilde k$ and $\tilde k^\prime$, while  no Markov Inequality is needed for its application. Most significantly, the security is controlled by a security parameter that can make the quantitative level arbitrarily small. There is {\it no} such security parameter in QKD. Thus, $d\leq \epsilon$ does {\it not} offer the equivalent of such ``semantic security" in QKD.

Composition security of the ``expansion'' kind under {\it d} follows from the the fact that $\parallel \rho_i-\sigma_i\parallel_1\leq \epsilon_i, i\in\overline{1-m}$,  implies

\begin{equation}
\parallel \rho_1\otimes \cdots \otimes \rho_m-\sigma_1\otimes\cdots \otimes\sigma_m\parallel_1 \leq \sum^m_{i=1}\epsilon_i
\end{equation}

\noindent In contrast to the case of interpretation (i) or (ii), this (11) does not guarantee KPA security which is a ``contraction".  Furthermore, there are problems in adopting (11) to the form (4) or (5), which is overcome \cite{9} by the incorrect interpretation (ii).

We have gone through all the security significance of $d$ that can be drawn from the literature, correct and incorrect ones. We will next derive valid operational guarantee from the security condition $d\leq \epsilon$.

\section{SEQUENCE ERROR PROBABILITY GUARANTEE FROM $d$}

Theoretical quantities such as $I_{ac}$ and {\it d} have no direct operational or empirical meaning in information theory already \cite{13}. They surely do not in cryptography, and the translation of {\it d} into operational probability guarantee is a central issue in the foundation of information-theoretic cryptography. A generalization of interpretation (iii) is

\begin{equation}
\mathcal P_r \leq \mathcal P_i + \epsilon^\prime(\epsilon)
\end{equation}

\noindent where $\epsilon^\prime$ is some function of $\epsilon$ for $d\leq \epsilon$. Note that although (6) is a stronger security claim than (12), it does not imply it became (12) holds with certainty. Under (12), empirical meanings of {\it d} are obtained when the performance $\mathcal{P}_r$'s are Eve's success probabilities of various kinds. Generally the probabilities we are concerned with are Eve's optimal probabilities   $p_1(\tilde K)$ of successfully estimating $\tilde K$, which is any possible subset of $K$ including $K$ itself, during key generation and in a KPA when $K$ is used.  We may want to bound such probabilities to some acceptable level $\epsilon(\tilde K)$ which, ideally, should be close to $2^{-|\tilde K|}$ where $|\tilde K|$ is the bit length of $\tilde K$,
                                                                                                                                                                       \begin{equation}
p_1(\tilde{K})\leq\epsilon(\tilde{K})
\end{equation}

\noindent When $\tilde K$ consists of $m$ bits, $m\leq n$, the problems involved are $M$-ary decisions, $M=2^m$, classically or quantum mechanically. Eve is going to utilize all her side information and measures on her probe to best estimate $\tilde K$. The usual optimal detection theory [11,14] concerns the $\tilde K$-averaged optimal probability only.

Such averaged performance seems hard to avoid and we would need Markov Inequality (MAI) for a random variable $X$ to convert it to an individual guarantee [15],
\begin{equation}
P_r[|X|\geq\delta]\leq\frac{E[|X|]}{\delta}
\end{equation}

\noindent  For the application at hand, MAI needs to be used twice to get $\epsilon^\prime$ from $\epsilon$, from averages over privacy amplification codes and either measurement result $y$ or individual $k$. Such individual guarantee is needed to rule out with a high probability any disastrous breach of security.  It is evident that the trace distance or variational distance corresponding to a small $p(k)$ or $p(y)$ can be large for any given $d$ level. (In contrast, under interpretation (i) or (ii) one only needs MAI once for privacy amplification.) Since typically $\epsilon \gg \frac{1}{N}$, one can treat $\epsilon$  as a probability and minimizes the total failure probability, the probability that the specified security level is not guaranteed. This is important especially in crypto security because one needs to take the users' worst care performance to {\it guarantee} a minimum security level.

Thus, given $E[|X|] = \epsilon$ one can guarantee from (14) the level $|X|<\sigma$ with a probability $\geq 1-\epsilon/\sigma$. This gives the {\it failure probability} (failure to guarantee $|X|<\sigma$) of probability $<\epsilon/\sigma$. The ``<'' or ``$\leq$'' is not significant here because the bounds may be approached arbitrarily closely. When $\sigma$ itself is a similar failure probability the total failure probability $P_f$, assuming independence of the two failure events,  satisfies $P_f \leq 1- (1-\sigma)(1-\epsilon/\sigma) = \sigma + \epsilon/\sigma -\epsilon$. The best guarantee is these obtained at $\sigma = \epsilon^{1/2}$ with resulting $P_f \sim 2\epsilon^{1/2}$ for $\epsilon \ll 1$. Similarly, if there are two random parameters so that MAI needs to be sued twice, the total failure probability is minimized at $\sigma_1=\sigma_2=\epsilon^{1/3}$ with resulting $P_f \sim 3\epsilon^{1/3}$.

When Eve makes a measurement on her probe with result $Y$, from Bayes rule $d\leq \epsilon$ becomes \cite{16}
\begin{equation}
\sum_y p(y)v(p(k|y), U(k)) \leq \epsilon
\end{equation}

\noindent In lieu of applying MAI to $K$-average, one may apply it to the $Y$-average form (15). The first major problem in raw security is the optimal probability that Eve can estimate the whole $K$ correctly. This problem is more than the guarantee from a single distribution which would be offered by a single variational distance. It is readily shown\cite{16} through bounding on each measured result $y$ in (15) that the $K$-averaged optimum error probability $\bar {p_1}(k)$ is bounded  as in (12),
\begin{equation}
\bar p_1(k)\leq \frac{1}{N}+ \epsilon^\prime(\epsilon)
\end{equation}
\noindent with $\epsilon^\prime(\epsilon) = 3\epsilon^{1/3}$ for two uses of MAI.

Similarly, Eve's optimal subset $\tilde K$ probabilities averaged over $\tilde K$, $\bar p_1(\tilde K)$, are to be determined from $2^{|\tilde K|}$-ary optimum quantum decision. For some $\epsilon^\prime (\epsilon)$, one needs to show from $d\leq \epsilon$
\begin{equation}
\bar p_1(\tilde K) \leq 2^{-|\tilde K|} +\epsilon^\prime (\epsilon)
\end{equation}

\noindent For KPA, one needs to show for possibly another $\epsilon^\prime (\epsilon)$,
\begin{equation}
\bar p_1(\tilde K_2|K_1=k_1) \leq 2^{-|\tilde K_2|} + \epsilon^\prime (\epsilon)
\end{equation}
\noindent when a portion $K_1$ of $K$ is known to be $k_1$ and a subset $\tilde K_2$ of the rest $K_2$ is to be estimated. It appears difficult if not impossible to obtain (17)-(18)  in general [16], with or without the use of MAI. This is not surprising, since some $\tilde K$ may be poorly protected under just a condition $d\leq \epsilon$ on the overall $K$. The average performance overt the rest of $K$ is hard to quantify with or without MAI, especially for a KPA. (The proper subset probabilities given in \cite{2,10} are not Eve's optimal.) Thus, Eve's sequence success probabilities on $\tilde K \subset K$ in both raw and KPA security are not quantified (yet at least) under the security condition $d\leq \epsilon$. Of course, (17)-(18) follow  with a high probability under interpretation (i) or (ii).

Note that a quantitative statement together with a general proof is lacking on what security guarantee is obtained from $d$ against information locking leaks.

\section{BIT ERROR RATE GUARANTEE FROM $d$}
It is possible that in a wrong sequence decision on  $\tilde K$  for any   $\tilde K < K$ , Eve may nevertheless get more than half of the bits in $K$ correctly even if the ``a priori" $p(\tilde k)$ as derived from $p(k)$ is uniform. This is the bit error rate (BER) vs sequence error probability issue in communications. Thus, in addition to the sequence error probability Eve's typically better BER needs to be bounded also from $d\leq \epsilon$.  Eve's BER, to be denoted by $p_b$ , is clearly one fundamental operational criterion when $K$ is used in one-time pad format. The sequence error probability (rate) is needed if $K$ is used as the seed key of a conventional cipher, which however has no KPA security independently of Eve's quantum probe. Note that $p_b$ is different from a single-bit $\tilde K$ success probability and is not known to be determined by $d$. Under interpretation (i) or (ii), $p_b=\frac{1}{2}$ the ideal value with probability $\ge 1-\epsilon$.

      The only known general low error bound on $p_b$  is Fano Inequality which can be applied to both sequence error rate [15] and BER [17]. For $d\leq \epsilon$ its use for sequence error rate yields a bound less good than (16) for typical parameters [16]. The BER is defined to be the per bit error probability

\begin{equation}
p_b \equiv P_b(K) = \frac{1}{n}\sum^n_{i=1}P_e(i)
\end{equation}

\noindent where   $P_e(i)$   is the probability that the $i$th bit in $K$ is incorrectly obtained from Eve's estimate of $K$. Fano Inequality gives in this case
\begin{equation}
n \mathcal H(p_b) \ge H(K)- I_{ac}
\end{equation}

\noindent where $\mathcal H(\cdot)$   is the binary entropy function, $H(K)$ the entropy of $K$ from $p(k)$. We can bound $H(K)$  in bits from $d\leq \epsilon$ using [15, theorem 17.3.3]
                                                                                                                                                             \begin{equation}
H(K) \ge n- \epsilon(n+\log{\frac{1}{\epsilon}}) \sim n(1-\epsilon)
\end{equation}
\noindent In (21) the term $\log{\frac{1}{\epsilon}}$ is usually small compared with $n$ while $I_{ac}$ is typically $\ll 1$ and can be neglected in (20). Thus (20)-(21) yields, for $p_b = \frac{1}{2}-\epsilon^\prime$  and small $\epsilon^\prime$, $\epsilon^\prime \leq (\epsilon/{4\log e})^{1/2}$. Since MAI needs to be used twice before this $\epsilon^\prime$ is applied, it is similar to the case of using it three times and the final $\epsilon^\prime$ is therefore

\begin{equation}
\epsilon^\prime   \leq \epsilon^{\frac{1}{4}}/2\sqrt{\log e}
\end{equation}

\noindent As expected, the BER guarantee (22) in the form (12) is worse than that of the corresponding sequence error.

Similar to the case of sequence error rate, it does not appear there is any readily derived $p_b(\tilde K)$ from (20) for raw and KPA security if Eve attacks proper subsets  $\tilde K \subset K$      via optimal $2^{|\tilde K|}$-ary quantum detection. From her specific attack and knowledge of the error correction and privacy amplification codes, Eve can estimate $p(k)$ and attack $\tilde K$ from such ``a priori'' information apart from her probe measurement. As also in the case of her sequence error probability, the best subset performance can be very good for her and it is difficult to bound her average subset performance.

\section{QUANTITATIVE GUARANTEE FROM $d$}

There has been the persistent "intuition" that a criterion should be fine if the level is brought down to a sufficiently small level, assuming the value zero in the ideal case. That seems to be true, but the whole question is how small is sufficiently small. It is a {\it quantitative} issue relative to the situation at hand. Even $I_{ac}$ gives security if it is smaller than $2^{-n}$ for an n-bit $K$ \cite{16}, which also follows from the $O(\log n/\epsilon)$ bits of $K$ needed in a KPA \cite{5} for large leak of deterministic bits.

Equality in (16) may be achieved [10]. Since $\epsilon^\prime$ gives the probability level that the entire $K$ can be found by Eve in raw security, it must be very small even if not close to $2^{-n}$. Similar $p_1\sim\epsilon^\prime$ is obtained for $I_{ac}/n \leq \epsilon$ \cite{10}.   In this regard, the necessity of using Markov Inequality for individual guarantee to eliminate disastrous security breach causes great difficulty in practice, from (22) and $\epsilon^\prime = \epsilon^{\frac{1}{3}}$ in (16).  The problem then is that it is very hard to get a low enough level of $d$ or $I_{ac}$ in practice.

It is rare to find an experimental QKD system with quantified security, but the one in [18] gives $I_{ac}\sim2^{-21}$ for $n\sim4000$.  Similar to (16), after two applications of MAI, the $p_1$ level of $2^{-21}$ becomes $\sim 2^{-7}$, a drop over four orders of magnitude.  For the six-state BB84 protocol theoretically analyzed recently \cite{19} with $d\leq \epsilon= 10^{-9}$ (and key rate $0.01 {~} bps$), $\bar p_1$ from (16)  becomes $10^{-3}$,  a drop of six orders of magnitude from fairly secure to not at all secure. For BER, $\epsilon^\prime \sim 2^{-9}$ and thus on average one more bit out of 500 bits of $K$ may turn out correct to Eve in addition to half of $K$. This can be compared to the binomial fluctuation level of one in $10^4-10^5$ bits for the $n$ in [19]. For the system of [18] it is possible to bound $d$ via $p_1$ [16] and $\epsilon^\prime$ in (22) may get up to 25\%, though for such large value the approximation of $\mathcal H(p_b)$ leading to (21) may not be accurate. However, it is clear that drastic breach of security is not adequately ruled out with the above numerical values. No subset $\tilde K$ or KPA probabilities of the form (17)-(18) can be numerically estimated, since it is not known whether they hold for what $\epsilon^\prime(\epsilon)$.

\section{Conclusion}

We have shown that the QKD strong security claim (i) or (ii) fails to obtain under the security guarantee $d\leq \epsilon$ while (iii) does not address the operational security levels. We provide operational security meaning to $d$ via bounds on Eve's sequence error probability and bit error rate when she attacks the whole key $K$, the guaranteed levels are far worse than $d$ itself and no result can yet be obtained on her optimal attacks on subsets of $K$ or on her known-plaintext attacks. This also causes fundamental problem on the final key rate, because better general security cannot be obtained  by further privacy amplification without settling these security criteria issues first.

 Similar to $I_{ac}$ the criterion $d$ gives some measure of security to a QKD protocol . However, it does not rule out huge leakage with appreciable probability, especially when the achievable level is inadequate for the necessary individual guarantee which seems to be the case in current as well as future practice. Indeed, even the formulation of an operationally meaningful security criterion for known-plaintext attacks does not yet exist.  In a future paper, we will show the actual security level cannot even be quantified due to other factors not properly taken into account in the security analysis thus far. It appears that either QKD should be treated like any other existing cryptosystem  with unknown fundamental security, or new approaches and especially new protocols need to be developed for fundamental security guarantee.

\section{Acknowledgment}

I would like to thank S. Abruzzo, G. Kanter, C. Portmann, and M. Raginsky for useful inputs. This work was supported by the Air Force Office of Scientific Research.

\end{document}